\documentclass[aps,prl,twocolumn,superscriptaddress,showpacs]{revtex4}
\usepackage{graphicx}
\usepackage{dcolumn}
\usepackage{bm}

\begin{document}

\title{Measurements of Angular Distributions of Muons From Upsilon Meson Decays\\
in \mbox{\boldmath{$p\overline p$}} Collisions at \mbox{\boldmath{$\sqrt{s}=1.96\;{\rm TeV}$}}}

\affiliation{Institute of Physics, Academia Sinica, Taipei, Taiwan 11529, Republic of China}
\affiliation{Argonne National Laboratory, Argonne, Illinois 60439, USA}
\affiliation{University of Athens, 157 71 Athens, Greece}
\affiliation{Institut de Fisica d'Altes Energies, ICREA, Universitat Autonoma de Barcelona, E-08193, Bellaterra (Barcelona), Spain}
\affiliation{Baylor University, Waco, Texas 76798, USA}
\affiliation{Istituto Nazionale di Fisica Nucleare Bologna, $^{ee}$University of Bologna, I-40127 Bologna, Italy}
\affiliation{University of California, Davis, Davis, California 95616, USA}
\affiliation{University of California, Los Angeles, Los Angeles, California 90024, USA}
\affiliation{Instituto de Fisica de Cantabria, CSIC-University of Cantabria, 39005 Santander, Spain}
\affiliation{Carnegie Mellon University, Pittsburgh, Pennsylvania 15213, USA}
\affiliation{Enrico Fermi Institute, University of Chicago, Chicago, Illinois 60637, USA}
\affiliation{Comenius University, 842 48 Bratislava, Slovakia; Institute of Experimental Physics, 040 01 Kosice, Slovakia}
\affiliation{Joint Institute for Nuclear Research, RU-141980 Dubna, Russia}
\affiliation{Duke University, Durham, North Carolina 27708, USA}
\affiliation{Fermi National Accelerator Laboratory, Batavia, Illinois 60510, USA}
\affiliation{University of Florida, Gainesville, Florida 32611, USA}
\affiliation{Laboratori Nazionali di Frascati, Istituto Nazionale di Fisica Nucleare, I-00044 Frascati, Italy}
\affiliation{University of Geneva, CH-1211 Geneva 4, Switzerland}
\affiliation{Glasgow University, Glasgow G12 8QQ, United Kingdom}
\affiliation{Harvard University, Cambridge, Massachusetts 02138, USA}
\affiliation{Division of High Energy Physics, Department of Physics, University of Helsinki and Helsinki Institute of Physics, FIN-00014, Helsinki, Finland}
\affiliation{University of Illinois, Urbana, Illinois 61801, USA}
\affiliation{The Johns Hopkins University, Baltimore, Maryland 21218, USA}
\affiliation{Institut f\"{u}r Experimentelle Kernphysik, Karlsruhe Institute of Technology, D-76131 Karlsruhe, Germany}
\affiliation{Center for High Energy Physics: Kyungpook National University, Daegu 702-701, Korea; Seoul National University, Seoul 151-742, Korea; Sungkyunkwan University, Suwon 440-746, Korea; Korea Institute of Science and Technology Information, Daejeon 305-806, Korea; Chonnam National University, Gwangju 500-757, Korea; Chonbuk National University, Jeonju 561-756, Korea}
\affiliation{Ernest Orlando Lawrence Berkeley National Laboratory, Berkeley, California 94720, USA}
\affiliation{University of Liverpool, Liverpool L69 7ZE, United Kingdom}
\affiliation{University College London, London WC1E 6BT, United Kingdom}
\affiliation{Centro de Investigaciones Energeticas Medioambientales y Tecnologicas, E-28040 Madrid, Spain}
\affiliation{Massachusetts Institute of Technology, Cambridge, Massachusetts 02139, USA}
\affiliation{Institute of Particle Physics: McGill University, Montr\'{e}al, Qu\'{e}bec, Canada H3A~2T8; Simon Fraser University, Burnaby, British Columbia, Canada V5A~1S6; University of Toronto, Toronto, Ontario, Canada M5S~1A7; and TRIUMF, Vancouver, British Columbia, Canada V6T~2A3}
\affiliation{University of Michigan, Ann Arbor, Michigan 48109, USA}
\affiliation{Michigan State University, East Lansing, Michigan 48824, USA}
\affiliation{Institution for Theoretical and Experimental Physics, ITEP, Moscow 117259, Russia}
\affiliation{University of New Mexico, Albuquerque, New Mexico 87131, USA}
\affiliation{The Ohio State University, Columbus, Ohio 43210, USA}
\affiliation{Okayama University, Okayama 700-8530, Japan}
\affiliation{Osaka City University, Osaka 588, Japan}
\affiliation{University of Oxford, Oxford OX1 3RH, United Kingdom}
\affiliation{Istituto Nazionale di Fisica Nucleare, Sezione di Padova-Trento, $^{ff}$University of Padova, I-35131 Padova, Italy}
\affiliation{University of Pennsylvania, Philadelphia, Pennsylvania 19104, USA}
\affiliation{Istituto Nazionale di Fisica Nucleare Pisa, $^{gg}$University of Pisa, $^{hh}$University of Siena and $^{ii}$Scuola Normale Superiore, I-56127 Pisa, Italy}
\affiliation{University of Pittsburgh, Pittsburgh, Pennsylvania 15260, USA}
\affiliation{Purdue University, West Lafayette, Indiana 47907, USA}
\affiliation{University of Rochester, Rochester, New York 14627, USA}
\affiliation{The Rockefeller University, New York, New York 10065, USA}
\affiliation{Istituto Nazionale di Fisica Nucleare, Sezione di Roma 1, $^{jj}$Sapienza Universit\`{a} di Roma, I-00185 Roma, Italy}
\affiliation{Rutgers University, Piscataway, New Jersey 08855, USA}
\affiliation{Texas A\&M University, College Station, Texas 77843, USA}
\affiliation{Istituto Nazionale di Fisica Nucleare Trieste/Udine, I-34100 Trieste, $^{kk}$University of Udine, I-33100 Udine, Italy}
\affiliation{University of Tsukuba, Tsukuba, Ibaraki 305, Japan}
\affiliation{Tufts University, Medford, Massachusetts 02155, USA}
\affiliation{University of Virginia, Charlottesville, Virginia 22906, USA}
\affiliation{Waseda University, Tokyo 169, Japan}
\affiliation{Wayne State University, Detroit, Michigan 48201, USA}
\affiliation{University of Wisconsin, Madison, Wisconsin 53706, USA}
\affiliation{Yale University, New Haven, Connecticut 06520, USA}

\author{T.~Aaltonen}
\affiliation{Division of High Energy Physics, Department of Physics, University of Helsinki and Helsinki Institute of Physics, FIN-00014, Helsinki, Finland}
\author{B.~\'{A}lvarez~Gonz\'{a}lez$^z$}
\affiliation{Instituto de Fisica de Cantabria, CSIC-University of Cantabria, 39005 Santander, Spain}
\author{S.~Amerio}
\affiliation{Istituto Nazionale di Fisica Nucleare, Sezione di Padova-Trento, $^{ff}$University of Padova, I-35131 Padova, Italy}
\author{D.~Amidei}
\affiliation{University of Michigan, Ann Arbor, Michigan 48109, USA}
\author{A.~Anastassov$^x$}
\affiliation{Fermi National Accelerator Laboratory, Batavia, Illinois 60510, USA}
\author{A.~Annovi}
\affiliation{Laboratori Nazionali di Frascati, Istituto Nazionale di Fisica Nucleare, I-00044 Frascati, Italy}
\author{J.~Antos}
\affiliation{Comenius University, 842 48 Bratislava, Slovakia; Institute of Experimental Physics, 040 01 Kosice, Slovakia}
\author{G.~Apollinari}
\affiliation{Fermi National Accelerator Laboratory, Batavia, Illinois 60510, USA}
\author{J.A.~Appel}
\affiliation{Fermi National Accelerator Laboratory, Batavia, Illinois 60510, USA}
\author{T.~Arisawa}
\affiliation{Waseda University, Tokyo 169, Japan}
% ...

\collaboration{CDF Collaboration\footnote{With visitors from
$^a$Istituto Nazionale di Fisica Nucleare, Sezione di Cagliari, 09042 Monserrato (Cagliari), Italy,
$^b$University of CA Irvine, Irvine, CA 92697, USA,
$^c$University of CA Santa Barbara, Santa Barbara, CA 93106, USA,
$^d$University of CA Santa Cruz, Santa Cruz, CA 95064, USA,
$^e$Institute of Physics, Academy of Sciences of the Czech Republic, Czech Republic,
$^f$CERN, CH-1211 Geneva, Switzerland,
$^g$Cornell University, Ithaca, NY 14853, USA,
$^h$University of Cyprus, Nicosia CY-1678, Cyprus,
$^i$Office of Science, U.S. Department of Energy, Washington, DC 20585, USA,
$^j$University College Dublin, Dublin 4, Ireland,
$^k$ETH, 8092 Zurich, Switzerland,
$^l$University of Fukui, Fukui City, Fukui Prefecture, Japan 910-0017,
$^m$Universidad Iberoamericana, Mexico D.F., Mexico,
$^n$University of Iowa, Iowa City, IA 52242, USA,
$^o$Kinki University, Higashi-Osaka City, Japan 577-8502,
$^p$Kansas State University, Manhattan, KS 66506, USA,
$^q$Korea University, Seoul, 136-713, Korea,
$^r$University of Manchester, Manchester M13 9PL, United Kingdom,
$^s$Queen Mary, University of London, London, E1 4NS, United Kingdom,
$^t$University of Melbourne, Victoria 3010, Australia,
$^u$Muons, Inc., Batavia, IL 60510, USA,
$^v$Nagasaki Institute of Applied Science, Nagasaki, Japan,
$^w$National Research Nuclear University, Moscow, Russia,
$^x$Northwestern University, Evanston, IL 60208, USA,
$^y$University of Notre Dame, Notre Dame, IN 46556, USA,
$^z$Universidad de Oviedo, E-33007 Oviedo, Spain,
$^{aa}$CNRS-IN2P3, Paris, F-75205 France,
$^{bb}$Texas Tech University, Lubbock, TX 79609, USA,
$^{cc}$Universidad Tecnica Federico Santa Maria, 110v Valparaiso, Chile,
$^{dd}$Yarmouk University, Irbid 211-63, Jordan,
}}
\noaffiliation

\date{\today}

\begin{abstract}
The angular distributions of muons from
$\Upsilon(1S,2S,3S)\rightarrow \mu^+\mu^-$
decays are measured using data from $p\overline p$ collisions at
$\sqrt{s}=1.96\;{\rm TeV}$ corresponding to an integrated luminosity of
$6.7\;{\rm fb}^{-1}$ and collected with the CDF~II detector at the
Fermilab Tevatron.  This analysis is the first to report the full
angular distributions as functions of transverse momentum $p_T$
for $\Upsilon$ mesons in both the Collins--Soper and $s$--channel
helicity frames.  This is also the first measurement of spin alignment
of $\Upsilon(3S)$ mesons.  Within the kinematic range of $\Upsilon$
rapidity $|y|<0.6$ and $p_T$ up to $40\;{\rm GeV}/c$, the angular
distributions are found to be nearly isotropic.
\end{abstract}


\maketitle
Heavy quarkonium production in hadron collisions provides critical tests
of quantum chromodynamics (QCD) because it involves both short-distance and
long-distance contributions.  The cross sections for direct charmonium and
$\Upsilon$ production measured at the Tevatron in $p\overline p$
collisions~\cite{bib-oldcdf} greatly exceeded the predictions from
leading-order ``color singlet'' models~\cite{bib-csm}.
Further measurements of spin alignment in $J/\psi$, $\psi'$~\cite{bib-cdfjpsi}
and $\Upsilon$ production~\cite{bib-upscdf} proved to be in dramatic
disagreement with
predictions of ``color octet'' models that were developed to explain the
cross section results~\cite{bib-nrqcd}.  While spin alignment can provide very
sensitive tests for QCD models, the measured elements of the spin density
matrix for the spin-1 $\Upsilon$ states depend critically on the choice of
coordinate frame.

Recently, it has been pointed out that improved experimental measurements
are needed to clarify this situation~\cite{bib-faccioli}.  In general, the
angular distribution of the $\mu^+$ in the rest frame of an
$\Upsilon\rightarrow \mu^+\mu^-$ decay can be written as~\cite{bib-cs}
\begin{equation}
\frac{dN}{d\Omega}\sim 1 + \lambda_\theta\cos^2\theta + \lambda_\varphi \sin^2\theta\cos 2\varphi + \lambda_{\theta\varphi}\sin 2\theta\cos\varphi,
\label{eqn-dndomega}
\end{equation}
in which $\theta$ is the polar angle measured with respect to a quantization
axis, and $\varphi$ is the azimuthal angle measured with respect to
the production plane containing the $\Upsilon$ and the beam axis.
The coefficients are directly related to the elements
of the spin-density matrix for the ensemble of $\Upsilon$ states
observed~\cite{bib-noman}.
Previous studies at hadron colliders have measured only 
$\lambda_\theta$ in the $s$--channel helicity frame, where the
quantization axis coincides with the direction of the $\Upsilon$ momentum,
and cannot be transformed into different coordinate frames without
additional information.
This precludes model-independent comparisons
of results obtained in different coordinate frames or experimental
environments.  Improving on this experimental
situation requires not only measuring all three coefficients, but also
carrying out these measurements in multiple coordinate frames.  This would
allow a comparison of quantities
such as $\tilde\lambda=(\lambda_\theta+3\lambda_\varphi)/(1-\lambda_\varphi)$,
which should have the same value in different coordinate frames~\cite{bib-rot}.
Such a test would provide an important demonstration that observations
have not been seriously biased due to poor determination of experimental
acceptance or subtraction of highly non-isotropic backgrounds.

In this Letter, we report on the first analysis of angular distributions of
muons from $\Upsilon(1S,2S,3S)\rightarrow \mu^+\mu^-$ decays produced in $p\overline p$
collisions carried out using this formalism: the distributions are quantified
in both the $s$--channel helicity frame and in the Collins--Soper frame,
which approximates, on average, the direction of the velocity of the colliding
beams.  It is also the first analysis to provide information on
the angular distributions of muons in decays of the $\Upsilon(3S)$ state
which is more likely to be produced directly, rather than as a decay product
of higher mass quarkonium states.

The $\Upsilon\rightarrow\mu^+\mu^-$ decays were
collected using the CDF~II detector, which reconstructs charged-particle tracks
and measures their momenta using a six-layer silicon
strip detector~\cite{bib-svx} and a large-volume drift
chamber~\cite{bib-cot}, both with approximate cylindrical geometry and
positioned in a 1.4~T solenoidal magnetic field.  The tracking
detectors are surrounded by calorimeters and three separate muon detector
subsystems.  The CMU system~\cite{bib-cmu} consists
of four layers of drift tubes that are located outside the hadron
calorimeter and cover the central range of pseudorapidity
$|\eta|<0.6$.  The CMP muon system~\cite{bib-cmp} comprises more layers
of drift tubes and scintillator placed behind additional steel
absorber material and covers approximately $|\eta|<0.4$.
Extended muon coverage in the forward region $0.6<|\eta|<1$ is provided by
the CMX subdetector which is also constructed from scintillator
and drift tubes.

A three-level online event selection system (trigger) is used to identify
events that contain oppositely charged dimuon candidates.
The level-1 trigger requires two tracks with $p_T>1.5\;{\rm GeV}/c$
to be identified in the tracking chamber and to be geometrically correlated
with activity in the CMU or CMX muon systems~\cite{bib-xft}.
The level-2 trigger
requires the muons to have opposite charge, and it requires that one of the
muons have $p_T>3\;{\rm GeV}/c$ and that it be detected in both the CMU
and CMP systems.
Events satisfying the level-2 trigger are passed to the level-3 trigger
system, which employs a version of the full event reconstruction software
optimized for speed.
The event selection used in the level-3 trigger requires the
presence of two oppositely charged muon candidates with invariant mass
in the range $8<m(\mu^+\mu^-)<12\;{\rm GeV}/c^2$.  It requires one to be
reconstructed in both CMU and CMP systems with $p_T>4\;{\rm GeV}/c$ and the
other to be reconstructed in either the CMU or CMX detectors with
$p_T>3\;{\rm GeV}/c$.  In this Letter, we refer to the trigger scenario
that selects two central muons as CC, and the scenario that selects one
central and one forward muon as CF.
From $6.7\;{\rm fb}^{-1}$ of integrated luminosity, the
combination of these triggers provides an event sample containing
approximately $550\,000$ $\Upsilon(1S)$, $150\,000$ $\Upsilon(2S)$,
and $76\,000$ $\Upsilon(3S)$ decays.

The criteria used to select dimuon candidates closely follow
those previously used in Ref.~\cite{bib-bsmumu}.  Muon candidates are
reconstructed from tracks in the drift chamber that extrapolate to a
track segment reconstructed in at least one of the muon detector systems.
Geometric restrictions are imposed on muon candidates to ensure that they
are contained in regions of the detector with well-measured trigger
and track reconstruction efficiencies.  Efficiencies for the level-1
trigger and for
these selection criteria are measured using the unbiased track in
$J/\psi\rightarrow\mu^+\mu^-$ decays that were recorded using a single-muon
trigger.  This analysis also makes use of information from the CMP muon
system in the level-2 trigger.  The efficiency for selecting such muons
is measured using samples of $J/\psi\rightarrow\mu^+\mu^-$
decays obtained using triggers that required information from only CMU or CMX.

\begin{figure}
\includegraphics[width=245pt]{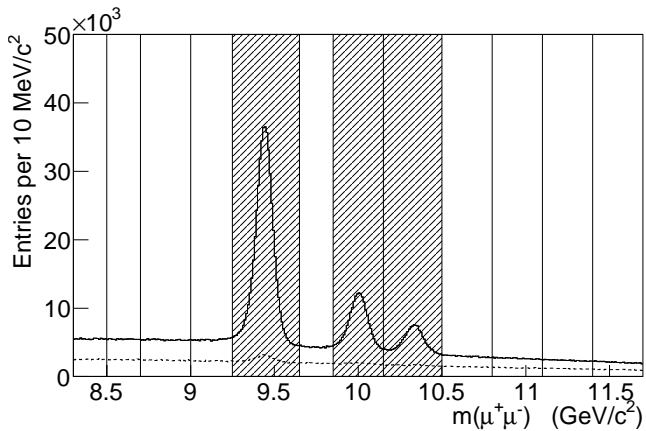}
\caption{\label{fig-mups}Mass distribution of $\mu^+\mu^-$ candidates
in the prompt (solid line) and displaced (dashed line) samples,
with the ranges of invariant mass used to select signal and sideband
events indicated.  Hatched regions indicate the three mass ranges containing
the $\Upsilon(nS)$ signals.}
\end{figure}

The angular distribution analysis is performed separately in each of the
12 ranges of dimuon mass shown in Fig.~\ref{fig-mups}.
The angular distributions of $\Upsilon$ decays are analyzed in eight
ranges of $p_T(\Upsilon)$ from $0$ to $40\;{\rm GeV}/c$ and are
restricted to the central region of rapidity $|y(\Upsilon)|<0.6$.
For a given range of transverse momenta, the sample of dimuon
candidates is divided into two subsamples according to whether one
of the muons is reconstructed precisely using measurements from the
silicon detector and its extrapolated trajectory misses the beam axis
by a distance $|d_0|>150\;{\rm \mu m}$.  Events with at least one muon
satisfying this requirement are referred to as the ``displaced'' sample
since they are consistent with the presence of a long-lived parent particle,
which is a characteristic feature of the dimuon background arising from
semileptonic decays of heavy quarks.
These criteria do not bias the angular distribution and, since the
displaced sample contains almost no $\Upsilon$ signal, it provides a
good description of the dimuon background that remains in the complementary
``prompt'' sample.  We verify that this is the case by comparing the
angular distributions of prompt and displaced samples
projected onto the $\cos\theta$ and $\varphi$ axes for mass ranges in the
sidebands of the $\Upsilon$ signals.  The level of agreement is quantified
by computing the Kolmogorov--Smirnov statistic for each distribution,
which suggests that any differences in their observed shapes are consistent
with statistical fluctuations.

The prompt sample contains most of the $\Upsilon\rightarrow\mu^+\mu^-$
signal, while a small fraction is retained in the displaced sample due
to the $30\;{\rm \mu m}$ resolution of the $d_0$ measurement
(see Fig.~\ref{fig-mups}).
The fraction $f_p$ that is present in the prompt sample is measured using a
simultaneous binned likelihood fit to the dimuon mass distributions of both
prompt and displaced samples.   The $\Upsilon$ signals are described by
Gaussian functions with common widths, their mass splittings constrained
to the known values~\cite{bib-pdg} and their yields
scaled by $f_p$ and $1-f_p$ in the prompt and displaced samples, respectively.
The value of $f_p$ ranges from 96-99\% depending on
the $\Upsilon$ $p_T$ and whether the candidate was recorded with the CC
or CF trigger scenario.

A second fit is performed to measure the mass-dependent ratio $r(m)$ of
the prompt and displaced mass distributions.  This is similar to the
first fit, but uses only the sidebands in the prompt sample,
$m(\mu^+\mu^-)<9\;{\rm GeV}/c^2$ or $m(\mu^+\mu^-)>10.5\;{\rm GeV}/c^2$,
to avoid the need to model the $\Upsilon(nS)$ line shapes.  The mass
distribution of the background in the displaced sample is parametrized
by a gamma function at low-$p_T$ and by an exponential function at higher
$p_T$.  The mass distribution of background in the prompt sample is
accurately described by the mass distribution in the displaced sample
multiplied by a scale factor that varies linearly with mass,
$r(m) = a+b m$, with the coefficients $a$ and $b$ determined from the fit.
The value of the function $r(m_{\Upsilon(1S)})$,
evaluated at the $\Upsilon(1S)$ mass, varies between 1.8--3.9 over the
range of dimuon $p_T$ considered.  In the subsequent analysis of angular
distributions, the value of this function and its uncertainty, both evaluated
at the center of each mass range containing the $\Upsilon$ signals, are used
to impose a Gaussian constraint on the background yield in the prompt sample.

\begin{figure}
\includegraphics[width=245pt]{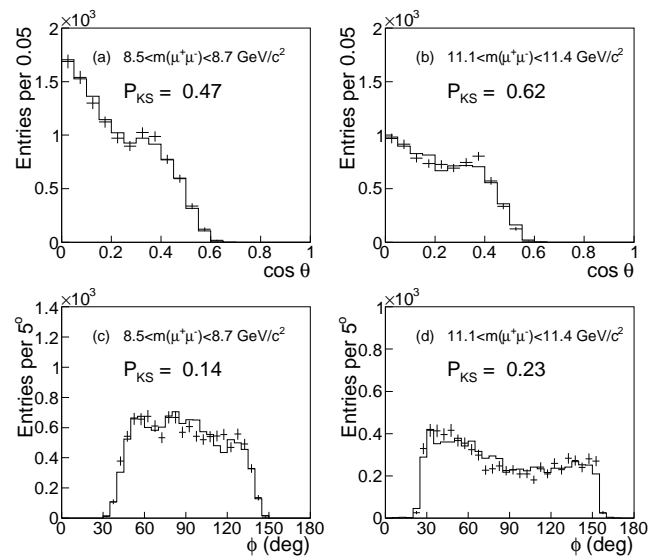}
\caption{\label{fig-angdisp}Comparisons of projected angular distributions
measured in the Collins--Soper frame for prompt (histograms) and
displaced (error bars) dimuon samples in the low-mass (a,c) and high-mass (b,d)
sidebands for events recorded with the CC trigger.  Displaced samples
are normalized to the number of events in the corresponding prompt samples.
The $p$-value of the Kolmogorov-Smirnov statistic comparing each
distribution is shown.}
\end{figure}

The displaced sample provides a good description of the angular distribution
of background in the prompt sample.  This observation is consistent with
all background arising from semileptonic decays of heavy quarks, or any
small non-heavy flavor background component having the same angular
distribution in prompt and displaced samples.
We observe good agreement between the angular distributions in the
prompt and displaced background, outside the $\Upsilon$
signal mass regions, even though the angular distributions change rapidly
with dimuon mass and $p_T$.  A typical example illustrating this
comparison is shown in Fig.~\ref{fig-angdisp}.
We then proceed to use the displaced sample to constrain the angular
distribution of the background when analyzing the angular distribution of
muons from $\Upsilon$ decays.

In mass ranges containing the $\Upsilon$ signals, we perform a third
simultaneous fit, in both the prompt and displaced samples, to the
distributions of
angles $(\cos\theta,\varphi)$ collected in $20\times 36$ discrete intervals.
Separate fits are performed for the two cases in which the angles represent
the direction of the positive muon with respect to the axes of the
$s$--channel helicity frame or those of the Collins--Soper frame.
For both frames, the angular distributions are described by linear
combinations of probability density functions for signal and background
components.  We factor these functions into an underlying
angular distribution, with that of the $\Upsilon$ signal parametrized using
Eq.~(\ref{eqn-dndomega}), and an acceptance function that accounts for the
geometry of the muon detectors and the kinematic restrictions imposed by the
trigger.  The parameters in the underlying angular distributions for the
$\Upsilon$ signal and for the background are determined using a simultaneous,
binned likelihood fit in which the expected numbers of events in each
discrete angular interval are expressed as
\begin{eqnarray}
  \frac{dN_p}{d\Omega_{ij}} &\sim&
   N_{\Upsilon} f_p {\cal A}_{\Upsilon}(\cos\theta_i,\varphi_j) \cdot
   w_\Upsilon(\cos\theta_i,\varphi_j;\vec\lambda_\Upsilon)+  \nonumber \\
   & & N_d s_p {\cal A}_{b}(\cos\theta_i,\varphi_j) \cdot
   w_{b}(\cos\theta_i,\varphi_j;\vec\lambda_{b}), \label{eqn-p} \\
  \frac{dN_d}{d\Omega_{ij}} &\sim&
   N_{\Upsilon} (1-f_p) {\cal A}_{\Upsilon}(\cos\theta_i,\varphi_j) \cdot
   w_\Upsilon(\cos\theta_i,\varphi_j;\vec\lambda_\Upsilon) + \nonumber \\
   & & N_d {\cal A}_{b}(\cos\theta_i,\varphi_j) \cdot
   w_b(\cos\theta_i,\varphi_j;\vec\lambda_{b}). \label{eqn-d}
\end{eqnarray}
In these expressions, $N_{\Upsilon}$ and $N_{d}$ are the $\Upsilon$ and
displaced background event yields, $f_p$ is the fraction of the $\Upsilon$
signal in the prompt sample, and $s_p$ is the ratio of the background yields
in the prompt and displaced samples, which is Gaussian-constrained to $r(m)$.
The acceptance for signal ${\cal A}_{\Upsilon}$ and background ${\cal A}_b$,
which are calculated separately, are described below.  
The modeling of the background angular distribution is improved by imposing
the additional kinematic restriction
$|p_T(\mu^+)-p_T(\mu^-)|<(p_T(\mu^+\mu^-)-0.5\;{\rm GeV}/c)$,
which removes back-to-back muons that have large
values of $\cos\theta$ in the $s$--channel helicity frame.
This has a negligible effect on the $\Upsilon(nS)$ acceptance
for $p_T(\Upsilon)>6\;{\rm GeV}/c$.  The angular distributions for $\Upsilon$
signal $w_\Upsilon$ and background $w_b$ are described by sets of parameters
$\vec\lambda_\Upsilon$ and $\vec\lambda_{b}$.
For the signal, $w_\Upsilon$ has the form of Eq.~(\ref{eqn-dndomega}),
whereas $w_{b}(\cos\theta,\varphi;\vec\lambda_b)$ has, in addition to the
terms in Eq.~(\ref{eqn-dndomega}), an empirical $\lambda_4\cos^4\theta$
term that parametrizes the background shape more accurately in some ranges of
$p_T$ and invariant mass.
The free parameters in the fit are
$N_\Upsilon$, $N_d$, $s_p$, $\vec\lambda_{\Upsilon}$, and
$\vec\lambda_b$, while $f_p$ is fixed to the value determined previously
from the first fit to the dimuon mass distributions.

The detector acceptance is calculated using a Monte Carlo (MC) simulation
in which dimuon events generated with isotropic distributions of decay
angles are processed using the standard CDF~II detector simulation and
event reconstruction programs.
Separate samples are simulated at fixed masses to calculate the acceptance
for the three $\Upsilon(nS)$ signals, while the acceptance for the dimuon
background is calculated using a continuum of dimuon invariant masses
ranging from $8$ to $12\;{\rm GeV}/c^2$.

\begin{figure}
\includegraphics[width=245pt]{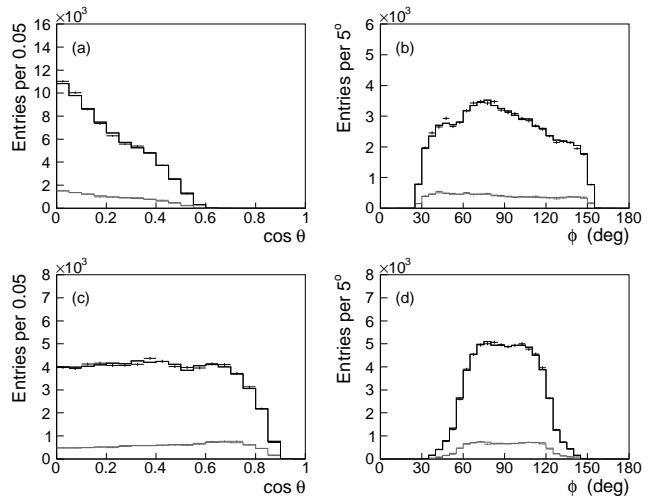}
\caption{\label{fig-1s}Examples of projections of angular variables
measured in the Collins--Soper (a,b) and the $s$--channel helicity (c,d)
frames for the range of invariant mass containing the $\Upsilon(1S)$ signal 
$4<p_T<6\;{\rm GeV}/c$ recorded with the CC trigger.  The prompt sample
is shown in black while the displaced sample is shown in grey.
The distributions from collision data are shown with error
bars and the results of the fits are indicated by the histograms.}
\end{figure}

Figure~\ref{fig-1s} shows distributions for the data and the best fit model
in one of the ranges of dimuon $p_T$ in the mass range containing
the $\Upsilon(1S)$ signal.
The shapes of the projected distributions are primarily determined by the
acceptance since the resolution with which $\cos\theta$ and $\varphi$ are
measured is much smaller than the binning chosen for the fit.  However,
the agreement between the data and the model in regions where the acceptance
changes rapidly, demonstrates that resolution effects are modeled accurately
in the detector simulation.
No significant discrepancy between the data and the fit model is observed
in $\chi^2$ tests applied to one dimensional projected distributions over
the analyzed range of dimuon mass and $p_T$.

Systematic uncertainties on the parameters $\lambda_\theta$, $\lambda_\varphi$
and $\lambda_{\theta\varphi}$ due to the limited precision
with which the trigger and reconstruction efficiencies are
determined are evaluated by repeating the analysis with acceptances
re-calculated with all efficiencies simultaneously varied by $\pm 1\sigma$.
The resulting change in the fitted parameters provides a
conservative estimate of the sensitivity to the measured acceptance.
Because the measured parameters
depend on the estimated background in the prompt sample, an alternate,
quadratic parametrization of the function $r(m)$ was also investigated.
The resulting small variations in the fitted angular distribution parameters
are treated as an additional systematic uncertainty.  Finally, the
contribution of the uncertainty in the fitted parameters due to the finite
MC sample size used to calculate the acceptance was estimated using
ensembles of MC simulations
with the same size used in the analysis of the
data.  The uncertainties due to finite MC sample size and the
determination of efficiencies are at most $30\%$ of the size of the statistical
uncertainty for the three $\Upsilon(nS)$ states, while the
uncertainty due to the treatment of the prompt scale factor function is
no more than $20\%$ of the statistical uncertainty.

\begin{figure}
\includegraphics[width=245pt]{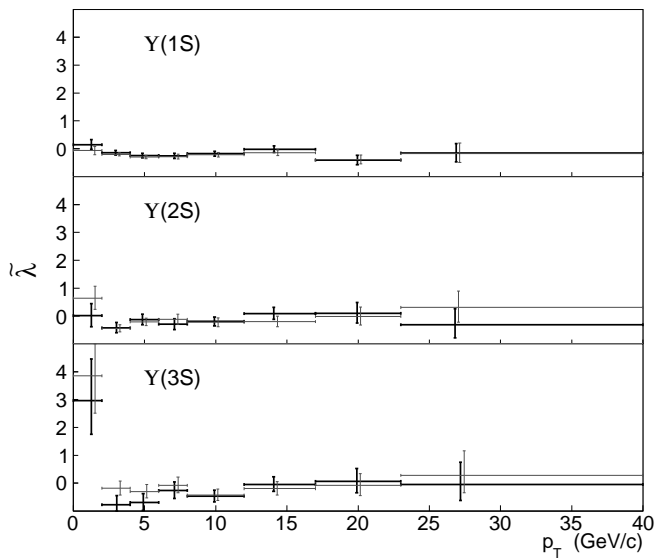}
\caption{\label{fig-lt}Rotational invariant $\tilde\lambda$
as a function of $p_T(\Upsilon)$ for the $\Upsilon(1S)$, $\Upsilon(2S)$, and
$\Upsilon(3S)$ states.  Values of $\tilde\lambda$ calculated in the
Collins--Soper frame are indicated by dark lines, while those calculated
in the $s$--channel helicity frame are indicated by grey lines and are
horizontally offset to slightly larger $p_{T}$ values for clarity.}
\end{figure}

Figure~\ref{fig-lt} shows the rotational invariant $\tilde\lambda$ which
is calculated from the measured values of $\lambda_\theta$ and
$\lambda_\varphi$ in each $p_T$ range for both the
Collins--Soper and $s$--channel helicity frames.
Uncertainties in $\tilde\lambda$ measured in the two coordinate frames are
highly correlated.  Monte Carlo simulations are used to calculate the expected
sizes of differences between the two values of $\tilde\lambda$ and in most
cases, the observed deviations are found to be consistent with purely
statistical fluctuations.  A systematic uncertainty derived from the difference
between $\tilde\lambda$ measured in the two coordinate frames is only
significant for the lowest three $p_T$ ranges of the $\Upsilon(3S)$.  In the
lowest $p_T$ range, the values of $\tilde\lambda$ measured for the
$\Upsilon(3S)$ differ by $2.4\;\sigma$, without accounting for systematic
uncertainties, and is the only case where the angular distribution is
observed to be significantly non-isotropic.  However, we can find no
evidence to suggest that this is due to a bias or systematic effect since
we do not see a similar trend for the $1S$ and $2S$ states and can find no
anomalous behavior in any of the underlying distributions.

\begin{figure}
\includegraphics[width=245pt]{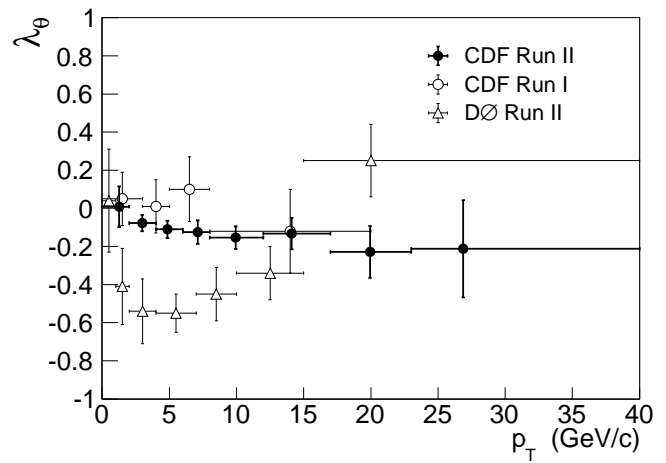}
\caption{\label{fig-cdfd01s}Comparison of the $\lambda_\theta$
parameter measured for $\Upsilon(1S)$ decays in the $s$--channel helicity
frame (solid symbols) with previous measurements from the
CDF~\cite{bib-upscdf} (open circles) and the D0~\cite{bib-upsd0}
(open triangles) experiments.}
\end{figure}

The values of $\tilde\lambda\approx 0$ suggest that the decays of all
three $\Upsilon(nS)$ resonances are consistent with an unpolarized mixture
of states.
Table~\ref{tab-alpha} lists the values of $\lambda_\theta$ measured in
the $s$--channel helicity frame for the $\Upsilon(1S)$, $\Upsilon(2S)$,
and $\Upsilon(3S)$ states, with the systematic uncertainties described above
added in quadrature~\cite{bib-url}.
\begin{table*}
  \begin{tabular}{crcrcr} \hline\hline
$p_T\;({\rm GeV}/c)$ & \multicolumn{1}{c}{$\lambda_\theta\left(\Upsilon(1S)\right)$} && \multicolumn{1}{c}{$\lambda_\theta\left(\Upsilon(2S)\right)$} && \multicolumn{1}{c}{$\lambda_\theta\left(\Upsilon(3S)\right)$} \\ \hline
0--2 & $0.01\pm 0.10\pm 0.05$ && $-0.07\pm 0.20\pm 0.05$  && $0.32\pm 0.36\pm 0.12$ \\[5pt]
2--4 & $-0.08\pm 0.04\pm 0.03$ && $-0.29\pm 0.10\pm 0.06$  && $-0.19\pm 0.17\pm 0.08$\\[5pt]
4--6 & $-0.11\pm 0.04\pm 0.03$ && $-0.11\pm 0.10\pm 0.05$ && $-0.01\pm 0.19\pm 0.09$ \\[5pt]
6--8 & $-0.12\pm 0.06\pm 0.04$ && $-0.09\pm 0.12\pm 0.07$ && $-0.48\pm 0.14\pm 0.13$ \\[5pt]
8--12 & $-0.15\pm 0.06\pm 0.02$ && $-0.18\pm 0.11\pm 0.03$ && $0.07\pm 0.18\pm 0.04$ \\[5pt]
12--17 & $-0.13\pm 0.08\pm 0.04$ && $-0.09\pm 0.15\pm 0.08$  && $0.14\pm 0.22\pm 0.07$  \\[5pt]
17--23 & $-0.23\pm 0.13\pm 0.08$ && $0.06\pm 0.25\pm 0.12$  && $0.14^{+0.36}_{-0.33}\pm 0.12$ \\[5pt]
23--40 & $-0.21\pm 0.24\pm 0.11$ && $0.00^{+0.39}_{-0.33}\pm 0.16$ && $0.14^{+0.50}_{-0.42}\pm 0.13$ \\[5pt]
\hline\hline
\end{tabular}
\caption{\label{tab-alpha}Values of $\lambda_\theta$, measured in the
$s$--channel helicity frame for each range of $p_T(\Upsilon)$.
The first uncertainty is statistical while the second uncertainty is
systematic and is described in the text.}
\end{table*}
Figure~\ref{fig-cdfd01s} shows a comparison of the $\lambda_\theta$
parameter, measured for the $\Upsilon(1S)$ state in the $s$--channel helicity
frame, with previous measurements.
The current result is found to be statistically consistent with
the previous measurement from CDF~\cite{bib-upscdf}, which was made for
$|y|<0.4$ at $\sqrt{s}=1.8\;{\rm TeV}$ rather than $|y|<0.6$ and
$\sqrt{s}=1.96\;{\rm TeV}$.  Restricting the current measurement to $|y|<0.4$
does not change the results appreciably.
The current $\Upsilon(1S)$ result is inconsistent with
the previous measurement from the D0~experiment~\cite{bib-upsd0} at the
level of $4.5\sigma$.

In conclusion, we have measured the angular distributions of muons from
$\Upsilon(1S)$, $\Upsilon(2S)$, and $\Upsilon(3S)$ decays with $|y|<0.6$ and
in several ranges of transverse momentum up to $40\;{\rm GeV}/c$.
We find that the decay-angle distributions of all three $\Upsilon(nS)$ states
are nearly isotropic, as was suggested by previous
measurements~\cite{bib-upscdf} in the case of the $\Upsilon(1S)$.
This is the first
measurement to simultaneously determine the three parameters needed to
fully quantify the angular distribution of
$\Upsilon(nS)\rightarrow \mu^+\mu^-$ decays.  This is also the first
analysis to present information on the angular distribution of
$\Upsilon(3S)$ mesons produced in high energy $p\overline p$ collisions.

%
%
We thank the Fermilab staff and the technical staffs of the participating
institutions for their vital contributions. This work was supported by the
U.S. Department of Energy and National Science Foundation; the Italian
Istituto Nazionale di Fisica Nucleare; the Ministry of Education, Culture,
Sports, Science and Technology of Japan; the Natural Sciences and Engineering
Research Council of Canada; the National Science Council of the Republic of
China; the Swiss National Science Foundation; the A.P. Sloan Foundation;
the Bundesministerium f\"ur Bildung und Forschung, Germany; the Korean
World Class University Program, the National Research Foundation of Korea;
the Science and Technology Facilities Council and the Royal Society, UK;
the Russian Foundation for Basic Research; the Ministerio de Ciencia e
Innovaci\'{o}n, and Programa Consolider-Ingenio 2010, Spain; the Slovak
R\&D Agency; the Academy of Finland; and the Australian Research Council (ARC).


\begin{thebibliography}{99}

\bibitem{bib-oldcdf}F. Abe {\it et al.} (CDF Collaboration), Phys. Rev. Lett. {\bf 79}, 572 (1997); {\bf 75}, 4358 (1995).

\bibitem{bib-csm}E. Braaten, M.A. Doncheski, S. Fleming and M.L. Mangano, Phys. Lett. {\bf B333}, 548 (1994); M. Cacciari and M. Greco, Phys. Rev. Lett. {\bf 73}, 1586 (1994).

\bibitem{bib-cdfjpsi}T. Affolder {\it et al.} (CDF Collaboration), Phys. Rev. Lett. {\bf 85}, 2886 (2000).

\bibitem{bib-upscdf}D. Acosta {\it et al.} (CDF Collaboration), Phys. Rev. Lett. {\bf 88}, 161802 (2002).

\bibitem{bib-nrqcd}P. Cho and A.K. Leibovich, Phys. Rev. D {\bf 53}, 150 (1996); {\bf 53}, 6203 (1996).

\bibitem{bib-faccioli}P. Faccioli, C. Louren\c{c}o, J. Seixas, and H. K. Wohri,
  Phys. Rev. Lett. {\bf 102}, 151802 (2009);
  Eur. Phys. J. C {\bf 69}, 657 (2010);
  P. Faccioli, C. Louren\c{c}o, and
  J. Seixas, Phys. Rev. Lett. {\bf 105}, 061601 (2010);
  Phys. Rev. D {\bf 81}, 111502(R) (2010).

\bibitem{bib-cs}J.C. Collins and D.E. Soper, Phys. Rev. D {\bf 16}, 2219 (1977).

\bibitem{bib-noman}M. Noman and S. D. Rindani, Phys. Rev. D {\bf 19}, 207 (1979).

\bibitem{bib-rot}Rotational invariance of $\tilde\lambda$ follows from the
invariance of the expression
$\sum_{n=-J}^J (-1)^n \langle \psi|n\rangle\langle -n|\psi\rangle$, in which
$|\psi\rangle$ is a general state vector for a particle of spin $J$ and
$|n\rangle$ are the standard $J_z$ basis eigenstates.

\bibitem{bib-svx}A. Sill {\it et al.}, Nucl. Instrum. Methods A {\bf 447}, 1 (2000); C.S. Hill, {\it ibid.} {\bf 530}, 1 (2004).

\bibitem{bib-cot}T. Affolder {\it et al.}, Nucl. Instrum. Methods A {\bf 526}, 249 (2004).

\bibitem{bib-cmu}G. Ascoli {\it et al.}, Nucl. Instrum. Methods A {\bf 268}, 33 (1988).

\bibitem{bib-cmp}A. Artikov {\it et al.}, Nucl. Instrum. Methods A {\bf 538}, 358 (2005).

\bibitem{bib-xft} E.~J.~Thomson {\it et al.}, IEEE Trans. Nucl. Sci. {\bf 49}, 1063 (2002).

\bibitem{bib-bsmumu}T. Aaltonen {\it et al.} (CDF Collaboration), Phys. Rev. Lett. {\bf 107}, 102001 (2011); {\bf 100}, 101802 (2008);
                    D. Acosta {\it et al.} (CDF Collaboration), {\it ibid.} {\bf 93}, 032001 (2004).

\bibitem{bib-pdg}K. Nakamura {\it et al.} (Particle Data Group), J. Phys. G {\bf 37}, 075021 (2010).

\bibitem{bib-url}See Supplemental Material at
   [URL will be inserted by publisher] for complete tables of results for
  $\lambda_\theta$, $\lambda_\varphi$, $\lambda_{\theta\varphi}$ and $\tilde\lambda$ in both the $s$--channel helicity frame and Collins--Soper frame for
the $\Upsilon(1S)$, $\Upsilon(2S)$, and $\Upsilon(3S)$.

\bibitem{bib-upsd0}V.M. Abazov {\it et al.} (D0~Collaboration), Phys. Rev. Lett. {\bf 101}, 182004 (2008).

\end{thebibliography}
\end{document}